\pdfoutput=1

\documentclass[11pt,a4paper]{article}
\usepackage{jheppub}
\usepackage{amsmath}
\usepackage{multicol,bbm}
\usepackage{enumerate}
\usepackage{bibentry}
\usepackage{epstopdf}

\usepackage{amssymb}

\def\be{\begin{equation}}
\def\ee{\end{equation}}
\def\ba{\begin{eqnarray}}
\def\ea{\end{eqnarray}}
\def\nl{\nonumber\\}

\def\a{\alpha}

\def\d{\delta}
\def\ad{{\dot\alpha}}

\def\l{\langle}

\def\CP1{\mathbb{CP}^1}
\def\SL2C{\mathrm{SL}(2,\mathbb{C})}

\def\Z2{\mathbb{Z}_2}

\def\su2{{SU(2)}}

\def\a{{\alpha}}

\def\[{\left[}
\def\]{\right]}

\def\l{\lambda}
\def\e{\epsilon}

\def\s{\sigma}
\def\a{\alpha}

\def\({\left(}
\def\){\right)}
\def\[{\left[}
\def\]{\right]}
\def\lan{\langle}
\def\ran{\rangle}
\def\<{\langle}
\def\>{\rangle}

\def\i2{\frac{i}{2}}

\def\cN{{\cal N}}

\def\2F1{\,_2{\rm F}_1}

\begin{document}


\title{{Connected formulas for amplitudes in standard model}}


\author[a,b]{Song He,}
\author[a,c]{Yong Zhang}
\affiliation[a]{CAS Key Laboratory of Theoretical Physics, Institute of Theoretical Physics, Chinese Academy of Sciences, Beijing 100190, China}
\affiliation[b]{School of Physical Sciences, University of Chinese Academy of Sciences, No.19A Yuquan Road, Beijing 100049, China} 
\affiliation[c]{Department of Physics, Beijing Normal University, Beijing 100875, China}
\emailAdd{songhe@itp.ac.cn, yongzhang.th@gmail.com} 

\date{\today}

\abstract{
Witten's twistor string theory has led to new representations of S-matrix in massless QFT as a single object, including Cachazo-He-Yuan formulas in general and connected formulas in four dimensions. As a first step towards more realistic processes of the standard model, we extend the construction to QCD tree amplitudes with massless quarks and those with a Higgs boson. For both cases, we find connected formulas in four dimensions for all multiplicities which are very similar to the one for Yang-Mills amplitudes. The formula for quark-gluon color-ordered amplitudes differs from the pure-gluon case only by a Jacobian factor that depends on flavors and orderings of the quarks. In the formula for Higgs plus multi-parton amplitudes, the massive Higgs boson is effectively described by two additional massless legs which do not appear in the Parke-Taylor factor. The latter also represents the first twistor-string/connected formula for form factors.
}

\maketitle 

\section{Introduction and summary of results}

The ability of computing scattering amplitudes in gauge theories is crucial for the discovery of new physics beyond the standard model of particle physics. Recent years have witnessed tremendous progress in calculating amplitudes of various processes both at tree and loop level. Furthermore, remarkable hidden structures of gauge-theory amplitudes have been discovered which point us towards a deeper understanding of the fundamental aspects of QFT. A great deal of the progress have been triggered by Witten's twistor-string theory of ${\cal N}=4$ super-Yang-Mills (SYM)~\cite{Witten:2003nn}.
The connected prescription further refined in~\cite{RSV} gives a closed formula for any $n$-point tree amplitude in ${\cal N}=4$ SYM as a localized integral over the moduli space of $n$-punctured Riemann spheres; similar connected formulas have been proposed for supergravity~\cite{Cachazo-Geyer-1206.6511, Cachazo-Skinner-1207.0741,Cachazo-Mason-Skinner-1207.4712} and then derived from new twistor string theories~\cite{Skinner-1301.0868}.

In view of these advances, it is very tempting to ask if one can derive such twistor-string/connected formulas for more realistic standard model processes, especially those in QCD. Obviously this has been achieved for gluon amplitudes, which are identical in QCD and in ${\cal N}=4$ SYM. However, to our best knowledge no such formulas are available for other important QCD amplitudes, such as those with quarks or the Higgs boson. At tree level of course these amplitudes have been computed using other techniques, and connected formulas do not seem to be more efficient for actual computations. Nevertheless, the goal we are after is to find a {\it closed formulas} for these QCD amplitudes with arbitrary multiplicity. Among other things, such formulas would open up a new direction of studying standard-model amplitudes from twistor-string point of view.

The other motivation comes from the efforts in extending the scope of theories naturally described by the so-called Cachazo-He-Yuan (CHY) formulation, which is a new representation for S-matrix of massless particles~\cite{Cachazo-He-Yuan-1306.6575, Cachazo-He-Yuan-1307.2199,Cachazo-He-Yuan-1309.0885}. It can be seen as a generalization of connected formulas to any spacetime dimensions and to a large variety of theories~\cite{Cachazo-He-Yuan-1409.8256, Cachazo-He-Yuan-1412.3479}. The formula is an integral over moduli space of Riemann spheres localized by the universal, theory-independent {\it scattering equations}, which were originally proposed in~\cite{Cachazo:2013iaa}. When written in terms of spinor-helicity variables in four dimensions, one can reproduce old and new connected formulas in these theories~\cite{SkinnerEYM, Cachazo-Zhang-1601.06305, He:2016vfi, Cachazo:2016njl}. It has been shown that CHY and connected formulas can be derived from ambitwistor string theories in ten and four dimensions, respectively~\cite{MS, 4d ambitwistor}. Despite its success, it remains an important open question how to obtain CHY representations of all-multiplicity amplitudes for a given QFT's, {\it i.e.} field contents and Lagrangian.  For example, to our best knowledge, no closed-formula CHY representation is known for QCD amplitudes with quarks and with the Higgs boson.

CHY representation for fermions have been studied in~\cite{Weinzierl:2014ava, Weinzierl}, where tree amplitudes have been used as input; explicit CHY formulas have been obtained using gluon-gluino correlators in superstring theory~\cite{MS, Bjerrum-Bohr:2014qwa}, and in particular a closed formula is known for one pair of gluinos and arbitrary number of gluons~\cite{HeOliver}. However, such formulas become very complicated for more pairs of gluinos, and more importantly, it is still far from QCD amplitudes where (anti-) quarks are in (anti-) fundamental representation. On the other hand, CHY formulas have been obtained for amplitudes with massive scalars and gauge bosons from dimension reduction~\cite{Naculich1,Naculich2}, but these are very different from the standard model amplitudes with Higgs or W, Z bosons. For example, one cannot get amplitudes with a single Higgs boson, since the mass of the latter does not come from an extra-dimension components of the momentum. Such amplitudes are equivalent to form factors where the Higgs momentum becomes that of an off-shell operator, and finding a formula for these amplitudes amounts to finding the first CHY representation for form factors.

Thus for both formal and practical purposes, it would be very intriguing to reproduce the correct coupling with quarks and to incorporate the massive Higgs boson (or off-shell form factor), in the CHY/twistor-string formulation. In this paper, we initiate this line of research, by writing down connected formulas for tree amplitudes with quarks and Higgs boson in the standard model. We find that the obstacles mentioned above can be circumvented, as long as we use four-dimensional scattering equations such that CHY takes the form of connected formulas as from various four-dimensional (ambi-)twistor string theories.  Such a representation also fully exploits the simplicity of spinor-helicity variables and provides a new way of computing these standard model amplitudes.

What is special in 4d is that the scattering equations naturally split into $n-3$ sectors~\cite{Cachazo:2013iaa},  labeled by $k=2,3,\ldots, n-2$, and for our consideration they coincide with the helicity sectors of amplitudes. More specifically, in this paper we use the four-dimensional scattering equations originally derived in~\cite{4d ambitwistor} from four-dimensional ambitwistor strings, which are completely equivalent to scattering equations in Witten's twistor string theory~\cite{He:2016vfi}. For sector $k$, it is convenient to split the $n$ particles into a set of $k$ particles we call $-$, and the complimentary set $+$ with $n{-}k$ particles, then the equations read
\be\label{eq}
\tilde\lambda^{\dot\alpha}_I\,-\,\sum_{i\,\in\,+} \frac{\tilde\lambda^{\dot\alpha}_i}{(I\,i)}\,=\,0\,,\quad I\in -\,;\quad \lambda^{\alpha}_i\,-\,\sum_{I\,\in\,-} \frac{\lambda^{\alpha}_I}{(i\,I)}\,=\,0\,,\quad i\in +\,.
\ee
Here $\lambda_a^{\alpha}, \tilde\lambda_a^{\dot\alpha}$ for $a=1,2,\ldots,n$ are spinors of $n$ external particles, and throughout the paper we use index $I,i$ for labels in the two sets $-,+$. The scale for the locations of punctures does matter in 4d, with the two components parameterized as $\sigma_a^\alpha = {1\over t_a}(1, \sigma_a)$, and the two-brackets are defined as $(a\,b):=(\sigma_a-\sigma_b)/(t_a t_b)$. The $\SL2C$ redundancy of CHY scattering equations has been extended to GL$(2,\mathbb{C})$, which can be used to fix four out of the $2n$ variables $\{\sigma, t\}$, and four of the $2n$ equations are also redundant since they simply impose momentum conservation. We refer to \cite{He:2016vfi} for the derivation of \eqref{eq} from CHY scattering equations and the equivalence to the original equations in~\cite{Witten:2003nn, RSV}.

A connected formula expresses tree amplitude in helicity-sector $k$ as an integral over the $2n-4$ variables localized on the solutions of the equations \eqref{eq}. The integrands differ for different theories, and results are known in (super--) Yang-Mills and gravity~\cite{4d ambitwistor} (equivalent to the original connected formulas in~\cite{RSV, Cachazo-Skinner-1207.0741}), effective field theories including supersymmetric DBI~\cite{He:2016vfi}, and (super--) Einstein-Yang-Mills theory~\cite{SkinnerEYM}.

Here we summarize the main results of the paper: we add two new connected formulas for two classes of amplitudes in the standard model. First, we present a formula for color-ordered $n$-point gluon-quark tree amplitudes in massless QCD:
\be\label{QCDgen}
\displaystyle
A_n^{g; q \bar{q}}=\int \frac{\prod_{a=1}^n\,d^2\sigma_a}{{\rm vol\,GL}(2)}~\frac{{\cal J}_{\rm ferm}(\{\sigma_{q,\bar{q}}\})}{(12)(23)\cdots (n1)}~\prod_{I\,\in\,-} \delta^2\left(\tilde\lambda_I-\sum_{i\,\in\,+} \frac{\tilde\lambda_i}{(I\,i)}\right)~\prod_{i\,\in\,+} \delta^2\left(\lambda_i-\sum_{I\,\in\,-} \frac{\lambda_I}{(i\,I)}\right)\,,
\ee
where $d^2 \sigma_a:=d \sigma_a d t_a/t_a^3$~\footnote{Details for modding out GL(2) redundancies and converting 4 redundant delta functions to give momentum conservation will be explained later. Throughout the paper $A_n=\delta^4 (P) M_n$ denotes the amplitudes with momentum-conserving delta functions.}, and the integral over $2n-4$ variables are localized by the $2n-4$ delta functions of scattering equations~\eqref{eq}; $-$ ($+$) denotes the set of gluons and quarks with negative- (positive-) helicities. As we will review shortly, \eqref{QCDgen} is almost identical to the connected formula for pure-gluon amplitudes, except for the presence of a Jacobian factor ${\cal J}_{\rm ferm}$. It is remarkable that this is a rational function of $\sigma$'s of the quarks and anti-quarks. We will work out the precise form of ${\cal J}_{\rm ferm}$, which depends on the flavors and helicities of the quark pairs, but independent of any information of the glouns.  

For example, for the pure-gluon case ${\cal J}=1$ by definition, and for one pair of quark-antiquark say, $\{I,i\}$, it is simply ${\cal J}_{I,i}=1/(I\,i)$ with the convention $I\in -, i\in +$. As we will discuss in sec. \ref{sec:QCD} , \eqref{QCDgen} can be derived from the fact that all gluon-quark amplitudes follow from linear combinations of gluon-gluino ones in ${\cal N}=4$ SYM~\cite{Dixon:2010ik, Schuster:2013aya}. In general it is clear how to construct ${\cal J}$ for any number of quark pairs, but it becomes more and more involved as the number increase. In this paper we explicitly write down compact form of ${\cal J}$ for up to four quark pairs, and verify that the result agrees with that in~\cite{Dixon:2010ik}.

In sec.~\ref{sec:Higgs} we study connected formulas for amplitudes with a Higgs plus multi-partons, where the coupling to Higgs is treated as an effective interaction vertex. The dominant contribution is from the top-quark loop which are integrated out in the large $m_t$ limit and result in a effective vertex, e.g. with gluons, of the form $\propto H~{\rm Tr}~G^{\mu\nu}~G_{\mu \nu}$. A observation in~\cite{Dixon:2004za, Badger:2004ty} is that the vertex can be written in terms of self-dual and anti-self-dual parts of the gluon fields $(\phi~{\rm Tr}~ G_{\rm sd}^{\mu \nu}~G_{{\rm sd},\,\mu \nu}~+~\phi^{\dagger}~{\rm Tr}~G_{\rm asd}^{\mu \nu}~G_{{\rm asd},\,\mu \nu})$, where the Higgs field decompose into $H=\phi+\phi^{\dagger}$. It is advantageous to study helicity amplitudes for $\phi$ plus $n$ partons and those for $\phi^+$ plus $n$ partons (related to each other by parity), and and the amplitudes for $H$ is the sum of the two $A^{n_g; H}_{n{+}1}=A^{n_g; \phi}_{n{+}1}+A^{n_g; \phi^\dagger}_{n{+}1}$.

It turns out that the key for writing down connected formula here is to assign two on-shell legs $x,y$ for the Higgs, with $k_H=\lambda_x \tilde\lambda_x+\lambda_y \tilde\lambda_y$ and they correspond to two additional punctures $\sigma_x$ and $\sigma_y$. For amplitudes with  $\phi$ ($\phi^\dagger$), we simply assign $x,y$ together with the $+$ ($-$ resp.) set for the scattering equations. A particularly nice form follows form fixing the four variables in $\sigma^\alpha_x$, $\sigma^\alpha_y$ using the  GL(2) redundancy:
\ba\label{Higgs}
\displaystyle
A^{n_g; \phi}_{n{+}1}&=&\lan x\,y\ran^2 \int  \frac{\prod_{a=1}^n\,d^2\sigma_a}{(12)(23)\cdots (n1)} \prod_{I\,\in\,-}\,\delta^2\big(\tilde\lambda_I-\sum_{i\,\in\,+,x,y} \frac{\tilde\lambda_i}{(I\,i)}\big)~\prod_{i\,\in\,+,x,y}\,\delta^2\big(\lambda_i-\sum_{I\,\in\,-} \frac{\lambda_I}{(i\,I)}\big)\,,\nl
A^{n_g; \phi^\dagger}_{n{+}1}&=&[ x\,y]^2 \int  \frac{\prod_{a=1}^n\,d^2\sigma_a}{(12)(23)\cdots (n1)} \prod_{I\,\in\,-, x,y}\,\delta^2\big(\tilde\lambda_I-\sum_{i\,\in\,+} \frac{\tilde\lambda_i}{(I\,i)}\big)~\prod_{i\,\in\,+}\,\delta^2\big(\lambda_i-\sum_{I\,\in\,-,x,y} \frac{\lambda_I}{(i\,I)}\big)\,,\nl
\ea
Note that in the formula, the ``Parke-Taylor factor" does not concern $x,y$ which makes sense as they ``bond" as a colorless scalar. The remarkable property of \eqref{Higgs} is that it does not depend on individual momenta of $x,y$ but only on their sum. In sec.~\ref{sec:Higgs}, we will discuss in details the motivation and consistency checks for \eqref{Higgs}, including various checks against known amplitudes and correct factorizations. Finally, by combining \eqref{QCDgen} and \eqref{Higgs} one obtains formulas for Higgs plus multi-parton amplitudes, with gluons and massless quarks.

\section{Connected formula for massless QCD amplitudes}\label{sec:QCD}

In this section we derive connected formula for massless QCD amplitudes, \eqref{QCDgen} with explicit form for the Jacobian ${\cal J}$ for up to four quark pairs. Following the idea of~\cite{Dixon:2010ik, Schuster:2013aya}, we write gluon-quark amplitudes as linear combinations of gluon-gluino amplitudes in ${\cal N}=4$ SYM. Before presenting results for ${\cal J}$, we first review the connected formula for SYM amplitudes. We emphasize that for pure-gluon amplitudes, these formulas are nothing but CHY formulas reduced to 4d~\cite{Cachazo:2013iaa, He:2016vfi}, but now it becomes natural to work in helicity sectors and to include supersymmetries.

Recall that with ${\cal N}=4$ supersymmetry, it is convenient to introduce the Grassmann-odd variable $\eta^A$ with $A=1,\cdots,4$ the SU(4) R-symmetry index, and the supermultiplet is combined to a on-shell superfield
\be
\Phi^{\mathrm{SYM}}(\eta)=g^++\eta^A\psi_A+\frac{1}{2!}\eta^A\eta^B\phi_{AB}+\frac{1}{3!}\eta^A\eta^B\eta^C\e_{ABCD}\bar{\psi}^D+\eta^1\eta^2\eta^3\eta^4 g^-\,,
\ee
where $g^{\pm}$, $\psi, \bar{\psi}$ and $\phi$ denote gluons, gluinos and scalars. The superamplitude is then a function of the on-shell superspace $\{\lambda^{\a}_a, \tilde\lambda^{\ad}_a, \eta^A_a\}$ for $a=1,2,\ldots,n$. As originally proposed by Witten~\cite{Witten:2003nn}, dependence on $\eta$ can be simply accounted by including fermionic delta functions analogous to those with $\tilde\l$'s, which gives the super-amplitudes in ${\cal N}\!=\!4$ SYM. This is completely parallel in the connected formula for ${\cal N}=4$ SYM from 4d ambitwistor strings~\cite{4d ambitwistor}, where $\eta$'s are included exactly as in the $\tilde\lambda$-half of \eqref{eq}:
 \begin{align}\label{SYM}
\mathcal{A}_{n,k}^{\cN=4}\,=\,\displaystyle \int \frac{\prod_{a=1}^n\,d^2\sigma_a}{{\rm vol\,GL}(2)}~& \prod_{I\in -}~\delta^2\left(\tilde\lambda_I-\sum_{i\in +} \frac{\tilde\lambda_i}{(I\,i)}\right)~\prod_{i\in +}~\delta^2\left(\lambda_i-\sum_{I\in -} \frac{\lambda_I}{(i\,I)}\right)\nl
& \times~\prod_{I\in -}~\delta^{0|4}\left(\eta_I-\sum_{i\in +} \frac{\eta_i}{(I\,i)}\right)~\frac{1}{(12)(23)\cdots (n1)}\,.
\end{align}
Note that there is $\mathrm{GL}(2,\mathbb{C})$ redundancy to be fixed in the measure: the most convenient way to do so is to delete $d^2\sigma_a\,d^2\sigma_b$ and compensate it with a factor $t_at_b(\s_a-\s_b)^2$. Similarly 4 (8) of the bosonic (fermonic) delta functions are redundant and can be pulled out for (super)-momentum conservation, i.e. $\delta^{4|8} \big( \sum_{a=1}^n\l_a^{\a} (\tilde{\l}_a^{\ad} | \eta_a^{A}) \big)$; e.g. if we choose to those delta functions corresponding to $I,J\in -$, we compensate with a factor  $\lan I\,J\ran^{2-{\cal N}}=\lan I\,J\ran^{-2}$.

As mentioned above, the main advantage of the scattering equations \eqref{eq} (as opposed to the original ones in~\cite{Witten:2003nn, RSV}), is that the corresponding formula \eqref{SYM} is particularly nice for extracting helicity (or component) amplitudes. We will always choose the set $\pm$ to contain positive- (negative-) helicity particles; for gluon amplitudes we have \eqref{SYM} with fermionic delta functions replaced by identity, which comes from the fermionic integral $\prod_{I\in -} d^4 \eta_I$.

Similarly it is straightforward to extract gluon-gluino amplitudes. For example, with one pair of gluinos $\{\bar{\psi}_{I,A}, \psi_i^A\}$ with $I\in-$, $i\in +$, after integrating out $\eta$'s for gluons, we are left with the fermionic integrals $(d^3 \eta_I)_A \, d \eta_i^A$ and the delta function $\delta^4 (\eta_I - \eta_i/(I\,i))$ which gives the Jacobian ${\cal J}=1/(I\,i)$. For amplitudes with $m$ pairs of gluinos  $(I_1, i_1), (I_2, i_2), \ldots, (I_m, i_m)$, the Jacobian after integrating $\eta$'s is exactly a determinant:
\be\label{gpsi}
A_{n,k}^{g;\psi\bar{\psi}}\,=\,\displaystyle \int \frac{\prod_{a=1}^n\,d^2\sigma_a}{{\rm vol\,GL}(2)}~\frac{\det M}{(12)(23)\cdots (n1)}~\prod_{I\in -}~\delta^2\left(\tilde\lambda_I-\sum_{i\in +} \frac{\tilde\lambda_i}{(I\,i)}\right)~\prod_{i\in +}~\delta^2\left(\lambda_i-\sum_{I\in -} \frac{\lambda_I}{(i\,I)}\right)\,,
\ee
where $M$ is a $m\times m$ matrix with element $M_{r\,s}=\frac{\d^{A_{I_r}\,A_{i_s}}}{(I_r\,i_s)}$ for $r,s=1,2,\ldots,m$. In defining the component amplitude, the ordering of fermionic integrals have been arranged such that the rows and columns correspond to $I_1,\ldots, I_m$ and $i_1,\ldots, i_m$. 

Now we are ready to work out Jacobians of \eqref{QCDgen} for gluon-quark amplitudes in massless QCD. The simplest case is when all quark lines are of the same flavor, where QCD amplitudes are identical to amplitudes in ${\cal N}=1$ SYM. This is because the color-ordered $g q \bar{q}$ vertex in QCD is identical to the $g \psi \bar{\psi}$ vertex in ${\cal N}=1$ SYM.  \eqref{gpsi}  directly holds for this case except all $\delta^{A_I\,A_i}=1$ since the gluinos only have one flavor. Thus the formula for single-flavor QCD amplitudes is given by \eqref{QCDgen} with ${\cal J}=\det M$ with $M_{r\,s}=\frac{1}{(I_r\,I_s)}$.

In general we consider $m$ quark lines all with distinct flavors, from which the case with some quark lines having same flavor can be constructed. As discussed in~\cite{Dixon:2010ik}, the key here is to choose linear combinations of ${\cal N}=4$ SYM amplitudes with various flavor-assignment of the gluinos, to avoid unwanted diagrams in ${\cal N}=4$ SYM (such as scalar-exchange or some gluon-exchange ones). It has been shown that such combinations can always be found for arbitrary $m$~\cite{Schuster:2013aya}, and we expect it to be the case for our construction as well. Here we will present the explicit form of ${\cal J}$ in \eqref{QCDgen} for all cases up to four quarks lines and confirm that the result agrees with that of \cite{Dixon:2010ik}. For all cases with $m>4$ that we have tested, our construction also works but the form of ${\cal J}$ becomes more and more complicated.

\begin{figure}[htbp]
  \centering
  \includegraphics[width=15cm]{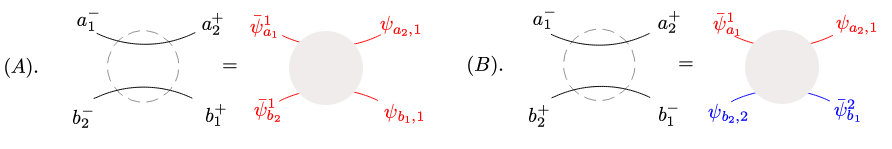}\\
  \caption{Two cases for amplitudes with 2 fermion pairs: QCD vs. SYM . }\label{fig1}
\end{figure}

We start with two pairs of quarks, and there are two orderings $(a_1^-,b_1^-,b_2^+,a_2^+)$ and $(a_1^-,a_2^+,b_1^-,b_2^+)$. Here we use $\{a_1, a_2\}$ and $\{b_1,b_2\}$ to denote the two quark lines with distinct flavors, and the two cases are referred to as splitting and alternating for the quark helicities; the remaining particles are gluons which can be put in arbitrary positions and do not affect ${\cal J}$. In the splitting case, we can identify them with the SYM amplitude where the flavors of the two gluino pairs are identical, i.e. $(\bar{\psi}_A, \bar{\psi}_A, \psi^A, \psi^A)$. The reason is that since a scalar-gluoino vertex e.g. $\phi_{AB} \psi^A \psi^B$ always change the flavor of the gluinos, this arrangement prevents the unwanted scalar exchanges between the two lines, and the helicities prevent unwanted gluon exchange to keep the quark flavors distinct (see fig.~\ref{fig1} A). Thus the Jacobian is the same as in the single-flavor case,
\ba
\mathcal{J}_{(a_1^-,b_1^-,b_2^+,a_2^+)}=\left|
\begin{array}{cc}
\frac{1}{(a_1a_2)}&\frac{1}{(a_1b_2)}\\
\frac{1}{(b_1a_2)}&\frac{1}{(b_1b_2)}
\end{array}
\right|\,.\label{split}
\ea
For the alternating case, no scalar exchange is allowed by the helicities, and to avoid unwanted gluon exchanges which would give QCD amplitudes with identical flavors, one has to use gluinos with two distinct flavors, i.e. $(\bar{\psi}_A, \psi^A, \bar{\psi}_B, \psi^B)$ for $B\neq A$ (see fig.~\ref{fig1} B). In the 2 by 2 matrix $M$ above, the off-diagonal entries vanish and the determinant reduces to the product of two diagonals,
\ba
\mathcal{J}_{(a_1^-,a_2^+,b_1^-,b_2^+)}=\frac{1}{(a_1a_2)}\frac{1}{(b_1b_2)}\,.\label{alter}
\ea

\begin{figure}
  \centering
  \includegraphics[width=15cm]{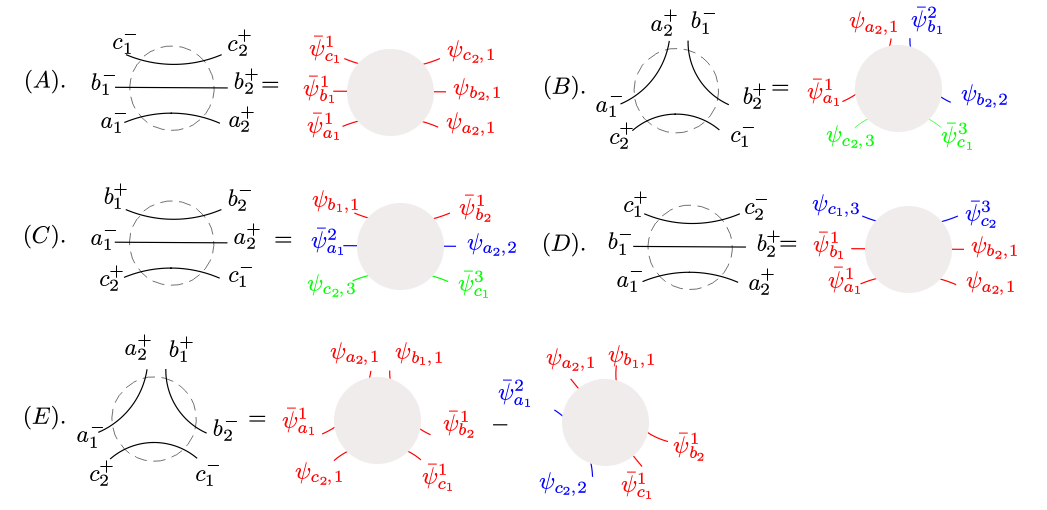}\\
  \caption{Five cases for amplitudes with 3 fermion pairs: QCD vs. SYM .}\label{fig2}
\end{figure}

Now we move to the case of three quark lines, where we have 5 distinct cases of quark orderings (other cases can be related to them by cyclicity or parity). For the splitting case, again the helicities prevent unwanted gluon exchanges and the result is the same as single-flavor case which prevents scalar exchanges (see fig.~\ref{fig2} A):
\ba
\mathcal{J}_{(a_1^-,b_1^-,c_1^-,c_2^+,b_2^+,a_2^+)}=
\left|
\begin{array}{ccc}
\frac{1}{(a_1a_2)}&\frac{1}{(a_1b_2)}&\frac{1}{(a_1c_2)}\\
\frac{1}{(b_1a_2)}&\frac{1}{(b_1b_2)}&\frac{1}{(b_1c_2)}\\
\frac{1}{(c_1a_2)}&\frac{1}{(c_1b_2)}&\frac{1}{(c_1c_2)}
\end{array}
\right|\,.\label{3.1}
\ea

On the other hand, there are two inequivalent alternating cases, $(a_1^-,a_2^+,b_1^-,b_2^+,c_1^-,c_2^+)$ and $(a_1^-,b_1^+,b_2^-,a_2^+,c_1^-,c_2^+)$. It is obvious that in both case, we can use gluinos with three distinct flavors to avoid unwanted gluon exchanges (see fig.~\ref{fig2} B,C); the 3 by 3 matrix $M$ becomes diagonal just like before , and we have
\ba\label{3.2}
{\cal J}_{(a_1^-,a_2^+,b_1^-,b_2^+,c_1^-,c_2^+)}\,=\,\mathcal{J}_{(a_1^-,b_1^+,b_2^-,a_2^+,c_1^-,c_2^+)}=\frac{1}{(a_1a_2)}\frac{1}{(b_1b_2)}\frac{1}{(c_1c_2)}\,.
\ea
Note that although the two Jacobians take the {\it same form} the two cases are different because they have different orderings. In other words, here the labesl $a_1, a_2,  b_1, b_2, c_1, c_2$ are different for the two cases, as is clear from fig.~\ref{fig2} B,C.

There are two more independent cases. For $(a_1^-,b_1^-,b_2^+,a_2^+,c_1^-,c_2^+)$, it is clear that we can assign same-flavor gluinos for lines $a$ and $b$ (a "sub-splitting" case) and we have to choose a different flavor for line $c$ since the helicities flipped between $a_2, c_1$, and between $c_2, a_1$ (see fig.~\ref{fig2} D). Thus all entries in the third column and row (corresponding to line $c$) vanish except for the diagonal, $1/(c_1\,c_2)$, and the Jacobian factorized as
\ba
\mathcal{J}_{(a_1^-,b_1^-,b_2^+,a_2^+,c_1^-,c_2^+)}=
\left|
\begin{array}{cc}
\frac{1}{(a_1b_2)}&\frac{1}{(a_1a_2)}\\
\frac{1}{(b_1b_2)}&\frac{1}{(b_1a_2)}
\end{array}
\right|
\frac{1}{(c_1c_2)}\,.\label{3.3}
\ea

The last case is $(a_1^-,a_2^+,b_1^+,b_2^-,c_1^-,c_2^+)$, which is slightly more complicated than cases above. Again one can assign same flavor gluinos for the three quark lines to  prohibit scalar exchanges. However, this single-flavor result contains the unwanted process where we have $(b_1, b_2)$, $(a_2, c_1)$ are two gluino pairs with same flavor and $(a_1, c_2)$ another pair with a different flavor~(see fig.\ref{fig2} E). This unwanted process is given by the Jacobian of the case~\ref{fig2} D, and we can subtract its contribution from the single-flavor result:
\ba
\mathcal{J}_{(a_1^-,a_2^+,b_1^+,b_2^-,c_1^-,c_2^+)}=
\left|
\begin{array}{ccc}
\frac{1}{(a_1a_2)}&\frac{1}{(a_1b_1)}&\frac{1}{(a_1c_2)}\\
\frac{1}{(b_2a_2)}&\frac{1}{(b_2b_1)}&\frac{1}{(b_2c_2)}\\
\frac{1}{(c_1a_2)}&\frac{1}{(c_1b_1)}&\frac{1}{(c_1c_2)}
\end{array}
\right|
-\frac{1}{(a_1c_2)}
\left|
\begin{array}{cc}
\frac{1}{(b_2a_2)}&\frac{1}{(b_2b_1)}\\
\frac{1}{(c_1a_2)}&\frac{1}{(c_1b_1)}
\end{array}
\right|\,.
\ea
This can be easily recognize as $\det M$ with the entry for $(a_1\, c_2)$ vanishes, $M_{a_1, c_2}=0$,
\ba
\mathcal{J}_{(a_1^-,a_2^+,b_1^+,b_2^-,c_1^-,c_2^+)}=
\left|
\begin{array}{ccc}
\frac{1}{(a_1a_2)}&\frac{1}{(a_1b_1)}&0\\
\frac{1}{(b_2a_2)}&\frac{1}{(b_2b_1)}&\frac{1}{(b_2c_2)}\\
\frac{1}{(c_1a_2)}&\frac{1}{(c_1b_1)}&\frac{1}{(c_1c_2)}
\end{array}
\right|\,.\label{3.4}
\ea

From these results we can infer the general rule for constructing ${\cal J}$ for any number of quark pairs: we observe that ${\cal J}$ is always given by a subset of terms from the determinant of the single-flavor matrix, $M_{I\,i}=\frac 1 {(I\,i)}$ where the rows and columns are given by quark labels in set $-$ and those in $+$, respectively. The subset is determined as follows: in the quark cyclic ordering, one inspects all adjacent labels that belong to two different quark lines; whenever they are of different helicities, one needs to remove terms in the expansion of $\det M$ that corresponding to this wrong contribution. In all but one cases explicitly presented in this paper, this can be done by simply setting some entries to zero. However, in certain cases first appeared for four quark-line case (see the end of the appendix~\ref{appendix}), we need to remove the contributions more carefully.

The validity of the rule have been checked thoroughly. For example, for alternating cases, \eqref{alter} and \eqref{3.2}, all quarks have flipped helicities compared to adjacent ones in the ordering, thus all off-diagonal entires are set to zero; similarly in \eqref{3.3}, the helicities of $c_1, c_2$ are flipped from those of $a_2, a_1$, while in \eqref{3.4} only that of $c_2$ is flipped from $a_1$, which explain the vanishing entries. In appendix~\ref{appendix}, we list all independent cases for four quark lines, where the form of ${\cal J}$ is always determined from the general rule. 

Naively one may conclude QCD amplitudes with more than four quark lines cannot be obtained in this way, since there are only four gluino flavors in ${\cal N}=4$ SYM. However, as shown in~\cite{Schuster:2013aya,Melia:2013epa}, this is not a problem as we can always reduce the number of gluinos by using a different (usually more complicated) combination. 

\begin{figure}[htbp]
  \centering
  \includegraphics[width=11cm]{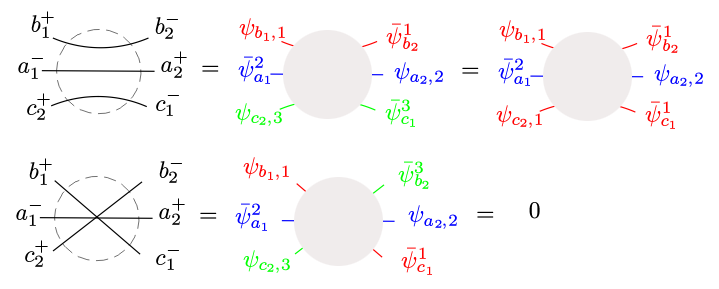}\\
  \caption{Equivalent forms with gluon flavors reduced, which implies a vanishing identity.}\label{fig3}
\end{figure}

We can already see how this works in some simple example, such as the second alternating case $(a_1^-,b_1^+,b_2^-,a_2^+,c_1^-,c_2^+)$. In \eqref{3.2} we used three gluino flavors but in fact we only need two since there can not be any gluon exchange between the two separated quark lines $b$ and $c$. Thus we can safely assign same gluino flavor for $b_1, b_2, c_1, c_2$, and obtain
\ba
\mathcal{J}_{(a_1^-,b_1^+,b_2^-,a_2^+,c_1^-,c_2^+)}=\frac{1}{(a_1a_2)}
\left|
\begin{array}{cc}
\frac{1}{(b_2 b_1)}&\frac{1}{(b_2 c_2)}\\
\frac{1}{(c_1 b_1)}&\frac{1}{(c_1 c_2)}
\end{array}
\right|\,.
\ea

This is equivalent to \eqref{3.2} (see fig.~\ref{fig3}), but we see that the number of flavors has been reduced by one. Since the two Jacobians give identical results, this also implies an interesting ``vanishing identity". The difference gives a vanishing amplitude which has, in addition to $(a_1, a_2)$, $(b_1, c_1)$ with one flavor and $(b_2, c_2)$ another flavor:
\begin{align}\label{vanishing}
{\footnotesize \int \frac{\prod_{a=1}^n\,d^2\sigma_a}{{\rm vol\,GL}(2)}~\frac{1/((a_1 a_2)\,(b_1c_1)\,(b_2 c_2))}{(12)(23)\cdots (n1)}~\prod_{I\in -}~\delta^2\left(\tilde\lambda_I-\sum_{i\in +} \frac{\tilde\lambda_i}{(I\,i)}\right)~\prod_{i\in +}~\delta^2\left(\lambda_i-\sum_{I\in -} \frac{\lambda_I}{(i\,I)}\right)=0\,.}
\end{align}

This is the connected-formula form of the fermion-crossing identity, see fig.~\ref{fig3}. In the simplest case of six-fermion (NMHV) amplitude, it can be shown using a residue theorem. After change of variables and Cauchy's theorem, the LHS of \eqref{vanishing} is equivalent to the tree contour of Grassmannian formula for $n=6, k=3$~\cite{ArkaniHamed:2009dn}, with an additional Jacobian from ${\cal J}_{\rm ferm}=1/((14)\,(25)\,(36))$. The Jacobian exactly cancels the three poles that defines the tree contour, which is why the contour integral vanishes.

Let's end with some remarks. CHY representation for QCD amplitudes has also been studied in~\cite{Weinzierl}. The major difference is that here we have an explicit formula for all multiplicities in four dimensions without assuming knowledge of any tree amplitudes. This again shows the remarkable simplicity of spinor-helicity and on-shell superspace in four dimensions, but it would also be very interesting to generalize to general dimensions and compare with~\cite{Weinzierl, HeOliver}. On the other hand, our derivation of QCD amplitudes from SYM ones is identical to that in~\cite{Dixon:2010ik}, but instead of using BCFW form of the amplitudes we have a compact, connected formula~\eqref{QCDgen}. In particular, we have seen that instead of combining different SYM amplitudes, all we need is to combine different SYM Jacobians to get a simple Jacobian ${\cal J}_{\rm ferm}$. In some sense, we have traded the complexities of the BCFW form for QCD helicity amplitudes~\cite{Dixon:2010ik} with the sum over solutions of \eqref{eq}.

\section{Higgs plus multi-parton amplitudes}\label{sec:Higgs}

In this section we turn to \eqref{Higgs} for Higgs plus multi-parton amplitudes, and we will first motivate it and then provide very strong consistency checks. Our connected formula is partly inspired by the relation between connected vs. disconnected prescription of twistor-string theory. The latter is known as CSW rules or MHV vertex expansion~\cite{Cachazo:2004kj}, which computes amplitudes with $k$ negative-helicity gluons as the sum of scalar Feynman diagrams with each vertex given by an off-shell continuation of MHV amplitudes:
\be\label{v1}
V(a,\ldots,J^-,\ldots, K^-,\ldots,b,q_{a,b})=\frac{\langle J\,K\rangle^4}{\langle q\,a \rangle\,\langle a\, a{+}1\rangle\,\cdots \langle b\,q\rangle}\,,\quad \lambda^\alpha_q:=q_{a,b}^{\alpha\,\dot\alpha} \tilde\mu_{\dot\alpha}\,,
\ee
where we have with two negative-helicity gluons, denoted by $J,K$, and the rest positive ones; $q$ is the off-shell leg for which we define $q_{r,s}:=k_r+k_{r{+}1}+\cdots +k_s$, and the CSW prescription involves a reference spinor $\tilde\mu$ to define the (holomorphic) spinor $\lambda_q$. It is interesting to see that the connected formula, can be viewed as an ``uplift"
of the MHV vertex: the helicity information is taken care of by scattering equations \eqref{eq}, and the integrand is given by the ``Parke-Taylor factor". Now for Higgs plus multi-gluon amplitudes, the CSW or disconnected formula is given in~\cite{Dixon:2004za}: $A^{n_g; \phi}_{n{+}1}$ is computed by scalar Feynman diagrams with MHV vertices of two types, the first without the Higgs field $\phi$ which is the same as \eqref{v1}, and the second type with $\phi$: $q_{r,s}:=k_r+k_{r{+}1}+\cdots +k_s$: %
\be\label{vertices}
V(\phi; a,\ldots,J^-,\ldots, K^-,\ldots,b,-q_{b{+}1,a{-}1})=\frac{\langle J\,K\rangle^4}{\langle q'\,a \rangle\,\langle a\, a{+}1\rangle\,\cdots \langle b\,q'\rangle}\,,
\ee
where we have defined $\lambda^\alpha_{q'}:=-q_{b{+}1,a{-}1}^{\alpha\,\dot\alpha}\,\tilde\mu_{\dot\alpha}$ with $-q_{b{+}1, a{-}1}=q_{a,b}+k_{\phi}$, and this vertex comes from the well-known MHV amplitudes with a Higgs field $\phi$:
\be\label{HiggsMHV}
A^{n_g;\phi}_{n{+}1}(\phi; 1,\ldots, J^-,\ldots, K^-,\ldots, n)=\delta^4(P)~\frac{\langle J\,K\rangle^4}{\langle 1\,2\rangle\,\langle 2\,3\rangle\,\cdots\,\langle n\,1\rangle}\,.
\ee

The massive momentum of the Higgs field $\phi$ can be written as the sum of two on-shell, massless momenta, $k_{\phi}=\lambda_x \tilde\lambda_x + \lambda_y \tilde\lambda_y$. More importantly, the structure of the CSW expansion for $A^{n_g; \phi}_{n{+}1}$ resembles that of a $(n{+}2)$-gluon amplitudes where instead of $\phi$ we have two additional {\it positive-helicity} gluons $x,y$. Thus for the kinematics (scattering equations) it is natural to assign $x,y$ together with the set $+$ ($-$ for $A^{n_g; \phi^\dagger}_{n{+}1}$).This can also be seen from the fact that the ``maximally googly"all-minus amplitude is non-vanishing: we need $x,y$ to be in the $+$ set for the scattering equations to have any solution.

On the other hand, we see that $x,y$ do not appear in the MHV vertex, so it is natural to have Parke-Taylor factor without $x,y$. This is expected as we use $x,y$ to represent the colorless scalar $\phi$, and the answer only depends on their total momentum $k_\phi$. To summarize, the first line of \eqref{Higgs} is the only possible ``uplift" of MHV vertices to a connected formula, which we record here for readers' convenience:
\be\label{Higgs1}
A^{n_g; \phi}_{n{+}1}=\lan x\,y\ran^2 \int  \frac{\prod_{a=1}^n\,d^2\sigma_a}{(12)(23)\cdots (n1)} \prod_{I\,\in\,-}\,\delta^2\left(\tilde\lambda_I-\sum_{i\,\in\,+,x,y} \frac{\tilde\lambda_i}{(I\,i)}\right)~\prod_{i\,\in\,+,x,y}\,\delta^2\left(\lambda_i-\sum_{I\,\in\,-} \frac{\lambda_I}{(i\,I)}\right)\,,
\ee
Note that we have already used GL$(2)$ redundancy to fix $\sigma_x, \sigma_y$ to arbitrary values, and if we recover the redundancy in the measure we need to insert the prefactor $1/(t_x t_y (\sigma_x-\sigma_y)^2)$. Together with $\lan x\,y\ran^2$, these factors are needed for the formula to have the correct mass dimension, little group and GL$(2)$ weight in $x,y$.

Now we provide strong consistency checks for the validity of \eqref{Higgs1}. Obviously it vanishes for $k=0,1$ since there is no solution to \eqref{eq}. Let's see how it reproduces the two simplest non-vanishing cases, namely MHV and all-minus amplitudes. To obtain the MHV amplitude \eqref{HiggsMHV}, it is convenient to pull out the four delta functions corresponding to $-=\{J,K\}$, to give momentum-conserving ones $\delta^4(P)$, which introduces a Jacobian $\lan J\,K \ran^2$. Then it is easy to see that the remaining $2n$ equations have a unique solution,
\be
{\rm MHV~solution}: \quad (a\,b)=\frac{\lan a\,b\ran}{\lan J\,K\ran}\,,\quad {\rm integral}~\to~\frac{\lan J\,K\ran^2}{\lan x\,y\ran^4 \lan 1\,2\ran\,\cdots \lan n\,1\ran}\,.
\ee
and combining with the prefactor it is obvious that we obtain \eqref{HiggsMHV}.  

The other extreme is the all-minus amplitude, which takes a particularly simple form~\cite{Dixon:2004za}
\be\label{allminus}
A(\phi; 1^-,\ldots, n^-)=\delta^4 (P)~\frac{m_H^4}{[1\,2]\,[2\,3]\,\cdots [n\,1]}\,,
\ee
where recall $P:=\sum_{a=x,y, 1}^n \lambda_a \tilde\lambda_a=0$ and $m_H^2=\lan x\,y\ran\,[x\,y]$. \eqref{Higgs1} is non-vanishing for $+=\emptyset$ exactly because $x,y$ serve as two positive-helicity ones which makes it similar to the anti-MHV case. Let's derive an equivalent form of \eqref{Higgs1} which manifestly has no dependence on $\lambda_x,\lambda_y$. We choose to pull out the delta functions for $\lambda_x, \lambda_y$ to impose momentum conservation. The Jacobian factor for doing this is $[x\,y]^2$ which combines with the prefactor $\lan x\,y\ran^2$ gives $m_H^4$, and we have
\begin{align}\label{Higgs2}
\displaystyle
A^{n_g; \phi}_{n{+}1}\,=\,\delta^4(P)\,m_H^4~\int  \frac{\prod_{a=1}^n\,d^2\sigma_a}{(12)(23)\cdots (n1)} \prod_{I\,\in\,-}\delta^2\left(\tilde\lambda_I-\sum_{i\,\in +, x,y} \frac{\tilde\lambda_i}{(I\,i)}\right)\prod_{i\,\in\,+}\delta^2\left(\lambda_i-\sum_{I\,\in\,-} \frac{\lambda_I}{(i\,I)}\right)\,.
\end{align}
For $+=\emptyset$, we use the $2n$ delta functions for $i=1,\ldots,n$ to get a unique solution again:
\be
{\rm anti-MHV~solution}: \quad (a\,b)=\frac{[a\,b]}{[x\,y]}\,,\quad {\rm integral}~\to~\frac{1}{[1\,2]\,[2\,3]\,\cdots\,[n\,1]}\,.
\ee 
and we see that the result is exactly \eqref{allminus}. Note that this result is not trivial from the disconnected (CSW) representation since it requires the sum over many MHV diagrams~\cite{Dixon:2004za}.

We also perform numerical checks for \eqref{Higgs2} (equivalent to \eqref{Higgs1}) to confirm that it gives correct results beyond the two extreme cases. We have evaluated \eqref{Higgs2} for all NMHV cases with $n=4,5,6$ and the $n=8$ NNMHV case. These are very non-trivial checks, since they correspond to NMHV amplitudes with 6,7,8 gluons and NNMHV with 8 gluons, where one sums over $4, 11,26$ and $66$ solutions respectively in the connected formula. It is satisfying to see that in all these cases \eqref{Higgs2} gives the same results as those from CSW rules~\cite{Dixon:2004za}.

More importantly, we have shown that \eqref{Higgs} has correct residues on all factorization poles, and here we only sketch the argument. It has been well established the connected formulas for ${\cal N}=4$ SYM (and pure-gluon) amplitudes have correct factorization limits, {\it c.f.}~\cite{Cachazo-Mason-Skinner-1207.4712}.  What we need here is the proof for the formula with $n+2$ gluons $1,2,\ldots, n, x,y$ with $\{x,y\}$ in all possible positions in the color ordering. Given the ordering, it is sufficient to consider $(k_1+k_2+\cdots+k_m)^2\to 0$, and one can show that in the Yang-Mills connected formula, both the measure (including delta functions) and the Parke-Taylor factor factorize nicely, and we have the correct factorization:
\be
M^{\rm YM}_{n{+}2} \to \sum_h M^{\rm YM}_{m{+}1} (1,2,\ldots, m, {\cal I}^h) \frac {1} {(k_1+\cdots+k_m)^2}  M^{\rm YM}_{n{-}m{+}3} (-{\cal I}^{-h}, m{+}1,\ldots,n; \{x,y\})\,,\nonumber
\ee
where $M$ is has momentum-conserving delta functions stripped, $A=\delta^4 (P)~M$, and the intermediate gluon ${\cal I}$ has momentum $k_{\cal I}=-(k_1+k_2+\cdots+k_m)$. Now we can use the same argument for $M^{n_g;\phi}_{n{+}1}$ with two slight modifications: one has to GL(2)-fix $\sigma_x, \sigma_y$ in the measure, and remove them in the Parke-Taylor factor. We have confirmed that both the measure and the Parke-Taylor factor still factorize as expected, except for the special case when $m=n$. This is the collinear limit of $x,y$ for$M^{\rm YM}_{n{+}2}$, and luckily it is not a possible factorization limit of $M^{n_g;\phi}_{n{+}1}$ as long as $m_H\neq 0$. Therefore, we have seen that \eqref{Higgs1} indeed has the correct behavior under general factorization limit \be\label{fac}
M^{n_g;\phi}_{n{+}1} \to \sum_h M^{\rm YM}_{m{+}1} (1,2,\ldots, m, {\cal I}^h) \frac {1} {(k_1+\cdots+k_m)^2}  M^{(n{-}m{+}1)_g; \phi}_{n{-}m{+}2} (-{\cal I}^{-h}, m{+}1,\ldots,n; \phi)\,,
\ee
with $m=2,\ldots, n{-}1$ (note that there is nothing special about the other collinear limit case $m=2$). Given \eqref{fac} the next step is to show that it has the desired large-$z$ behavior under a BCFW/CSW shift~\cite{BCF, BCFW}, which would give a complete proof of \eqref{Higgs1}. 

Last but not least, since fermions do not enter the effective vertex but only interact through gluons, we can combine \eqref{Higgs} with \eqref{QCDgen} to obtain Higgs plus multi-patron amplitudes. We have explicitly checked in various cases that our formula agrees with results in~\cite{Badger:2004ty}, including MHV and non-MHV cases for up to six partons and two quark lines.

\section{Discussions}

One of the most remarkable advances triggered by Witten's twistor string theory is a new formulation for S-matrix in massless QFT as a single object. This representation, known as CHY formulas in general and connected formulas in 4d, can be derived from localized worldsheet integrals or string correlators, in various (ambi-)twistor string theories. In this paper, we extend the construction to more realistic processes of the standard model, including QCD amplitudes with quarks and the Higgs boson. The new connected formulas are based on essentially the gluon scattering equations \eqref{eq}, with a Jacobian factor for the quarks in \eqref{QCDgen}, and the Higgs momentum shared by two special on-shell legs \eqref{Higgs}. The results are surprisingly compact and it is intriguing to see the coupling with quarks and the Higgs is naturally incorporated in this new representation. 

It is highly desirable to generalize our results to CHY formulation in general dimensions, which can shed new lights into how QFT interactions, in particular the Higgs mechanism, emerge from CHY/twistor string formulas. It would be interesting to extend our construction to other standard model process, such as trees entering in subleading-color loop amplitudes and those involving a electroweak vector boson (photon, W or Z). These amplitudes have very similar structures~\cite{Bern:2004ba,Badger:2005jv}, and we expect to have nice connected formulas for them as well. Furthermore, CHY formulation usually makes manifest nice properties of the amplitudes, such as color/kinematics duality and Bern-Carrasco-Johansson (BCJ) relations~\cite{Bern:2008qj}. Note that \eqref{QCDgen} is for color-ordered QCD amplitudes, and by QCD color-decomposition of~\cite{Johansson:2015oia} one immediately gets the full, color-dressed amplitude. The authors of~\cite{Johansson:2015oia} have shown that color-kinematics duality imply BCJ relations for QCD partial amplitudes (see also~\cite{delaCruz:2015dpa}), which we now see directly from \eqref{QCDgen}. In~\cite{Cachazo-1206.5970}, it has been shown that on the support of scattering equations, Parke-Taylor factors satisfy fundamental BCJ relations; the same argument works with \eqref{QCDgen} for fundamental BCJ with gluons, but not for those with quarks, due to the presence of ${\cal J}_{\rm ferm}$ which depends on quark orderings. Another interesting question is to study the result from double-copy of amplitudes with fermions such as in QCD (see \cite{delaCruz:2016wbr} for a recent exploration). 

Our result for Higgs plus gluon amplitudes, \eqref{Higgs}, is also a connected formula for the form factor with operator ${\rm Tr}~F^2$. This opens up a new direction of CHY/connected formulas for form factors and even more off-shell quantities such as correlation functions. It is straightforward to extend \eqref{Higgs} is to the form factor with chiral stress-tensor multiplet operators in ${\cal N}=4$ SYM, including ${\rm Tr}~\phi^2$ up to the chiral Lagrangian ${\cal L}$~\cite{Brandhuber:2011tv}. A more non-trivial and interesting question is how to obtain connected formulas for form factors with general operators in ${\cal N}=4$ SYM. In the limit that the momentum of the operator becomes soft, it gives tree amplitudes with one insertion of higher-dimensional operator, such as ${\rm Tr}~F^m$, which are of great interests for studying effective theory beyond QCD (c.f. \cite{Dixon:2004za}). For example,  for all effective vertices from the $\alpha'$-expansion of superstring theory such as  $F^4$ operator, it is straightforward to obtain their CHY and connected formulas since they are linear combination of YM amplitudes~\cite{Mafra:2011nv}. However, it remains an open question for operators that have no superstring origin, such as $F^3$~\cite{Huang:2016tag}.

Another important direction is CHY/twistor-string formulas at loop level, which has been studied for (super)-Yang-Mills and gravity amplitudes at one loop~\cite{ACS,Geyer-Mason-Monteiro-Tourkine-1507.00321,He:2015yua,Geyer-Mason-Monteiro-Tourkine-1511.06315,Cachazo:2015aol}. It is plausible that four-dimensional connected formulas can be generalized to loop level for QCD amplitudes/form factors and especially in ${\cal N}=4$ SYM. It is also highly desirable to study connections with BCFW/CSW representation, Grassmannian and on-shell diagrams~\cite{ArkaniHamed:2012nw}. In general we can use residue theorems to rewrite our connected formulas into ``disconnected" representations, {\it i.e.} sum of rational building blocks, which are more efficient for actual computations. One way of doing this~\cite{Arkani-Hamed-Bourjaily-Cachazo-Trnka-0912.4912, Dolan:2009wf, He-1207.4064} leads to Grassmannian contour integrals, whose residues are BCFW terms or on-shell diagrams. Thus our formulas imply Grassmannian formulas for QCD amplitudes and form factors (see~\cite{Frassek:2015rka} for a similar proposal), as well as other ``disconnected" representations to be further explored. We hope to systematically address these interesting questions in the future. 

\section*{Acknowledgments}
S.H.~thanks F.~Cachazo, C.S.~Lam and Zhengwen Liu for discussions. 

\appendix
\section{$\mathcal{J}_{\rm ferm}$ for four quark lines}\label{appendix}
Here we present ${\cal J}_{\rm ferm}$ of \eqref{QCDgen} for all independent cases with four quark lines.

\ba
&&\mathcal{J}_{(a_1^-,a_2^+,b_1^-,b_2^+,c_1^-,c_2^+,d_1^-,d_2^+)}=\mathcal{J}_{(a_1^-,a_2^+,b_1^-,c_1^+,c_2^-,d_1^+,d_2^-,b_2^+)}
=\mathcal{J}_{(a_1^-,a_2^+,b_1^-,c_1^+,d_1^-,d_2^+,c_2^-,b_2^+)}\nl
&=&\frac{1}{(a_1a_2)}\frac{1}{(b_1b_2)}\frac{1}{(c_1c_2)}\frac{1}{(d_1d_2)}\,.
\ea
\ba
&&\mathcal{J}_{(a_1^-,b_1^-,b_2^+,a_2^+,c_1^-,c_2^+,d_1^-,d_2^+)}=
\mathcal{J}_{(a_1^-,b_1^-,b_2^+,a_2^+,c_1^-,d_1^+,d_2^-,c_2^+)}=
\mathcal{J}_{(a_1^-,b_1^-,c_1^+,c_2^-,b_2^+,a_2^+,d_1^-,d_2^+)}\nl
&=&\left|
\begin{array}{cc}
\frac{1}{(a_1a_2)}&\frac{1}{(a_1b_2)}\\
\frac{1}{(b_1a_2)}&\frac{1}{(b_1b_2)}
\end{array}
\right|
\frac{1}{(c_1c_2)}\frac{1}{(d_1d_2)}\,.
\ea
\ba
&&\mathcal{J}_{(a_1^-,a_2^+,b_1^+,b_2^-,c_1^-,c_2^+,d_1^-,d_2^+)}=
\mathcal{J}_{(a_1^-,a_2^+,b_1^+,d_1^-,d_2^+,b_2^-,c_1^-,c_2^+)}=
\mathcal{J}_{(a_1^-,a_2^+,b_1^+,b_2^-,c_1^-,d_1^+,d_2^-,c_2^+)}\nl
&=&\left|
\begin{array}{ccc}
\frac{1}{(a_1a_2)}&\frac{1}{(a_1b_1)}&0\\
\frac{1}{(b_2a_2)}&\frac{1}{(b_2b_1)}&\frac{1}{(b_2c_2)}\\
\frac{1}{(c_1a_2)}&\frac{1}{(c_1b_1)}&\frac{1}{(c_1c_2)}
\end{array}
\right|
\frac{1}{(d_1d_2)}\,.
\ea
{\small \ba
\mathcal{J}_{(a_1^-,b_1^-,c_1^-,d_1^-,d_2^+,c_2^+,b_2^+,a_2^+)}=2\mathcal{J}_{(a_1^-,a_2^+,b_1^+,c_1^+,c_2^-,b_2^-,d_1^+,d_2^-)}=
\left|
\begin{array}{cccc}
\frac{1}{(a_1a_2)}&\frac{1}{(a_1b_2)}&\frac{1}{(a_1c_2)}&\frac{1}{(a_1d_2)}\\
\frac{1}{(b_1a_2)}&\frac{1}{(b_1b_2)}&\frac{1}{(b_1c_2)}&\frac{1}{(b_1d_2)}\\
\frac{1}{(c_1a_2)}&\frac{1}{(c_1b_2)}&\frac{1}{(c_1c_2)}&\frac{1}{(c_1d_2)}\\
\frac{1}{(d_1a_2)}&\frac{1}{(d_1b_2)}&\frac{1}{(d_1c_2)}&\frac{1}{(d_1d_2)}
\end{array}
\right|\,.
\ea}
\ba
\mathcal{J}_{(a_1^-,b_1^-,c_1^-,c_2^+,b_2^+,a_2^+,d_1^-,d_2^+)}=
\left|
\begin{array}{ccc}
\frac{1}{(a_1a_2)}&\frac{1}{(a_1b_2)}&\frac{1}{(a_1c_2)}\\
\frac{1}{(b_1a_2)}&\frac{1}{(b_1b_2)}&\frac{1}{(b_1c_2)}\\
\frac{1}{(c_1a_2)}&\frac{1}{(c_1b_2)}&\frac{1}{(c_1c_2)}
\end{array}
\right|
\frac{1}{(d_1d_2)}\,.
\ea
\ba\label{4fermion}
\mathcal{J}_{(a_1^-,b_1^-,b_2^+,a_2^+,c_1^-,d_1^-,d_2^+,c_2^+)}=
\left|
\begin{array}{cc}
\frac{1}{(a_1a_2)}&\frac{1}{(a_1b_2)}\\
\frac{1}{(b_1a_2)}&\frac{1}{(b_1b_2)}
\end{array}
\right|
\left|
\begin{array}{cc}
\frac{1}{(c_1c_2)}&\frac{1}{(c_1d_2)}\\
\frac{1}{(d_1c_2)}&\frac{1}{(d_1d_2)}
\end{array}
\right|\,.
\ea

\ba
\mathcal{J}_{(a_1^-,a_2^+,b_1^+,c_1^+,c_2^-,b_2^-,d_1^-,d_2^+)}=
\left|
\begin{array}{cccc}
\frac{1}{(a_1a_2)}&\frac{1}{(a_1b_1)}&\frac{1}{(a_1c_1)}&0\\
\frac{1}{(c_2a_2)}&\frac{1}{(c_2b_1)}&\frac{1}{(c_2c_1)}&\frac{1}{(c_2d_2)}\\
\frac{1}{(b_2a_2)}&\frac{1}{(b_2b_1)}&\frac{1}{(b_2c_1)}&\frac{1}{(b_2d_2)}\\
\frac{1}{(d_1a_2)}&\frac{1}{(d_1b_1)}&\frac{1}{(d_1c_1)}&\frac{1}{(d_1d_2)}
\end{array}
\right|\,.
\ea
\ba
\mathcal{J}_{(a_1^-,a_2^+,b_1^-,b_2^+,c_1^+,c_2^-,d_1^+,d_2^-)}=
\left|
\begin{array}{cccc}
\frac{1}{(a_1a_2)}&\frac{1}{(a_1b_2)}&\frac{1}{(a_1c_1)}&\frac{1}{(a_1d_1)}\\
0&\frac{1}{(b_1b_2)}&\frac{1}{(b_1c_1)}&\frac{1}{(b_1d_1)}\\
\frac{1}{(c_2a_2)}&\frac{1}{(c_2b_2)}&\frac{1}{(c_2c_1)}&0\\
\frac{1}{(d_2a_2)}&\frac{1}{(d_2b_2)}&\frac{1}{(d_2c_1)}&\frac{1}{(d_2d_1)}
\end{array}
\right|\,.
\ea
\ba
\mathcal{J}_{(a_1^-,a_2^+,b_1^+,b_2^-,c_1^-,c_2^+,d_1^+,d_2^-)}=&&
\left|
\begin{array}{cccc}
\frac{1}{(a_1a_2)}&\frac{1}{(a_1b_1)}&\frac{1}{(a_1c_2)}&\frac{1}{(a_1d_1)}\\
\frac{1}{(b_2a_2)}&\frac{1}{(b_2b_1)}&\frac{1}{(b_2c_2)}&\frac{1}{(b_2d_1)}\\
\frac{1}{(c_1a_2)}&\frac{1}{(c_1b_1)}&\frac{1}{(c_1c_2)}&\frac{1}{(c_1d_1)}\\
\frac{1}{(d_2a_2)}&\frac{1}{(d_2b_1)}&\frac{1}{(d_2c_2)}&\frac{1}{(d_2d_1)}
\end{array}
\right|
-
\left|
\begin{array}{cc}
\frac{1}{(a_1a_2)}&\frac{1}{(a_1b_1)}\\
\frac{1}{(d_2a_2)}&\frac{1}{(d_2b_1)}
\end{array}
\right|
\left|
\begin{array}{cc}
\frac{1}{(b_2d_1)}&\frac{1}{(b_2c_2)}\\
\frac{1}{(c_1d_1)}&\frac{1}{(c_1c_2)}
\end{array}
\right|\nl
&-&\left|
\begin{array}{cccc}
\frac{1}{(a_1c_2)}&\frac{1}{(a_1d_1)}\\
\frac{1}{(d_2c_2)}&\frac{1}{(d_2d_1)}
\end{array}
\right|
\left|
\begin{array}{cccc}
\frac{1}{(b_2b_1)}&\frac{1}{(b_2a_2)}\\
\frac{1}{(c_1b_1)}&\frac{1}{(c_1a_2)}
\end{array}
\right|\,.
\ea
The last case requires more than setting entries to zero. The single-flavor result contains two unwanted processes where we have either $(b1,b2)$, $(a2,c1)$ with same flavor, $(a1,c2)$ and $(d1,d2)$ with another same flavor,  or $(a1,a2)$, $(b1,d2)$ with same flavor and $(b2,d1)$,$(c1,c2)$ with another same flavor. The two unwanted processes are given by Jacobians of the case \eqref{4fermion}, and we need to subtract their contributions from the full determinant.

\end{document}